\documentclass[a4paper,11pt]{article}
\pdfoutput=1

\usepackage{jcappub}
\usepackage[T1]{fontenc}
\usepackage[utf8]{inputenc}
\usepackage{lmodern}

\usepackage{tikz}
\usepackage{booktabs}
\usepackage{pgfplots}
\pgfplotsset{compat=1.18}

\graphicspath{{./figs/}}

\definecolor{valecol}{rgb}{0,0.5, 1.}

\newcommand{\bm}[1]{\boldsymbol{#1}}

\newcommand{\nhat}{\bm{\hat{n}}}
\newcommand{\nhatp}{\bm{\hat{n}}^{\prime}}

\newcommand\T{\rule{0pt}{2.6ex}}       

\arxivnumber{2403.14580}

\title{\boldmath Tomographic redshift dipole:\\Testing the cosmological principle}

\author[a]{Pedro da Silveira Ferreira}
\author[b,c,d]{and Valerio Marra}

\affiliation[a]{PPGCosmo, Universidade Federal do Espírito Santo, 29075-910, Vitória, ES, Brazil}
\affiliation[b]{Dep.\ de Física, Universidade Federal do Espírito Santo, 29075-910, Vitória, ES, Brazil}
\affiliation[c]{INAF -- Osservatorio Astronomico di Trieste, 34131 Trieste, Italy}
\affiliation[d]{IFPU -- Institute for Fundamental Physics of the Universe, 34151, Trieste, Italy }

\emailAdd{dasferreira.pedro@gmail.com}
\emailAdd{valerio.marra@me.com}

\abstract{The cosmological principle posits that the universe is statistically homogeneous and isotropic on large scales, implying all matter shares the same rest frame. This principle suggests that velocity estimates of our motion from various sources should agree with the cosmic microwave background (CMB) dipole's inferred velocity of 370 km/s. Yet, for over two decades, analyses of radio galaxy and quasar catalogs have found velocities at odds with the CMB dipole, with tensions up to 5$\sigma$. In a blind analysis of BOSS and eBOSS spectroscopic data from galaxies and quasars across $0.2<z<2.2$,
we applied a novel dipole estimator for a tomographic approach, robustly correcting biases and quantifying uncertainties with realistic mock catalogs.
Our findings with eBOSS data ($0.6<z<2.2$), indicating a velocity of $196^{+92}_{-79}$ km/s, demonstrate a $2\sigma$ agreement with the CMB dipole when considering the full 3D vector distribution and a 3-to-6$\sigma$ tension with previous number count studies. This result supports the cosmological principle, emphasizing the consistency of our motion with the CMB across vast cosmic distances.
On the other hand, the BOSS data revealed potential unmodeled systematics; the estimator could not be minimized using the LOWZ set ($0.2<z<0.4$), and the CMASS set ($0.4<z<0.6$) presented results that pointed towards the southern hemisphere, conflicting with the CMB dipole.
Addressing the disparities with earlier number count analyses and understanding possible systematics in spectroscopic measurements will be essential to further validate the cosmological principle.}

\keywords{redshift surveys, cosmic flows, cosmological parameters from CMBR}

\begin{document}
\maketitle
\flushbottom

\section{Introduction}

The Cosmic Microwave Background (CMB) provides a snapshot of the early universe, with its temperature anisotropies offering insights into cosmic structures and dynamics. The most prominent of these anisotropies is the dipole, approximately 100 times larger than the fluctuations seen across smaller scales.
This dipole is traditionally attributed entirely to our proper motion relative to the CMB rest frame. Such interpretation leads to an inferred velocity $v=(369.82 \pm 0.11)$ km/s in the direction $(l,b)=(264.021\pm0.011,48.253\pm0.005)^\circ$~\cite{Akrami:2018vks}, which is used in astronomy to convert observed redshifts into cosmological ones~\cite{Peterson:2021hel}.

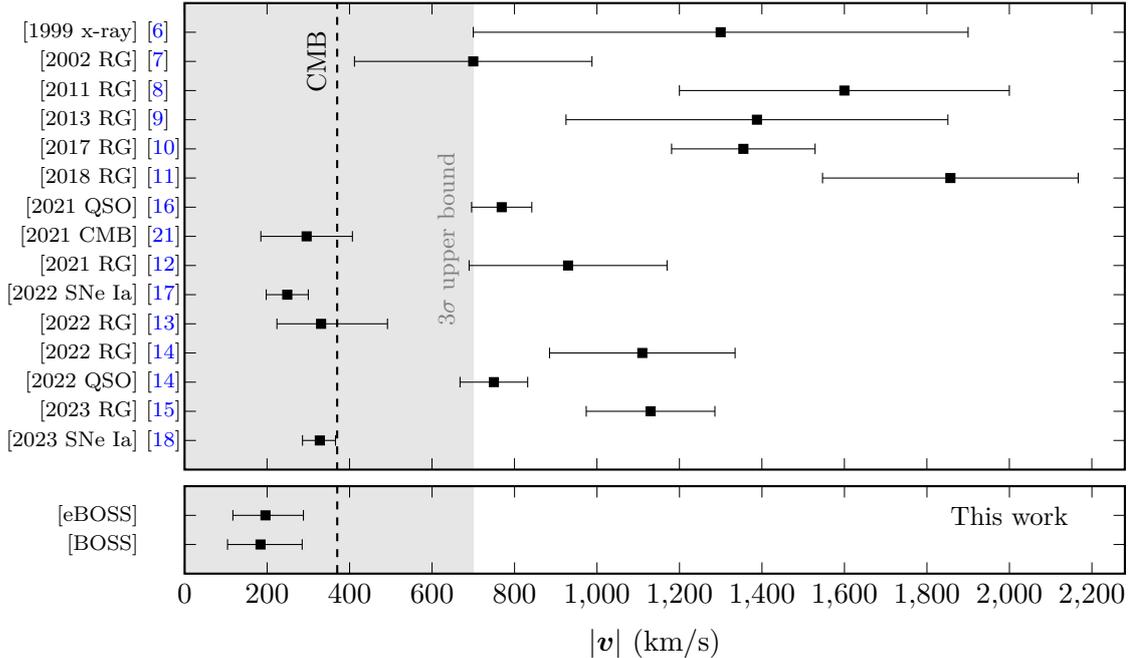
\begin{figure}[t]
\centering
\begin{tikzpicture}
    \begin{axis}[
        xmin=0,
        xmax=2280, 
        ymin=20, ymax=180, 
        line width=0.8pt,
        xlabel={},
        xticklabels={},
        ytick={170,160,150,140,130,120,110,100,90,80,70,60,50,40,30},
        yticklabels={
        \scriptsize{[1999 x-ray]}~\cite{Plionis:1998xj}$\text{\hspace{0.48mm}}$, 
        \scriptsize{[2002 RG]}~\cite{Blake:2002gx}$\text{\hspace{0.48mm}}$, 
        \scriptsize{[2011 RG]}~\cite{singaldip}$\text{\hspace{0.48mm}}$, 
        \scriptsize{[2013 RG]}~\cite{Rubart:2013tx}$\text{\hspace{0.48mm}}$,
        \scriptsize{[2017 RG]}~\cite{Colin:2017juj}$\!\!$, 
        \scriptsize{[2018 RG]}~\cite{Bengaly:2017slg}$\!\!$, 
        \scriptsize{[2021 QSO]}~\cite{Secrest:2020has}$\!\!$,
        \scriptsize{[2021 CMB]}~\cite{Ferreira:2020aqa}$\!\!$,
        \scriptsize{[2021 RG]}~\cite{Murray:2021frz}$\!\!$, 
        \scriptsize{[2022 SNe Ia]}~\cite{Horstmann:2021jjg}$\!\!$, 
        \scriptsize{[2022 RG]}~\cite{Darling:2022jxt}$\!\!$,
        \scriptsize{[2022 RG]}~\cite{Secrest:2022uvx}$\!\!$, 
        \scriptsize{[2022 QSO]}~\cite{Secrest:2022uvx}$\!\!$, 
        \scriptsize{[2023 RG]}~\cite{Wagenveld:2023kvi}$\!\!$,
        \scriptsize{$\text{\hspace{3.075mm}}$[2023 SNe Ia]}~\cite{Sorrenti:2022zat}$\!\!$}, 
        scale only axis=true,
        width=0.8\textwidth,
        height=0.4\textwidth,
        tick style={color=black},
        axis on top=true
    ]
    \fill[gray!20] (axis cs:0,-5.4) rectangle (axis cs:700,195.5);
    
    \addplot+[black, only marks, every mark/.append style={solid, fill=black,mark size= 1.5pt}, mark=square*,error bars/.cd, x dir=both,x explicit]
 coordinates { 
(1300,170)+-(600,600) 
(700,160)+-(288,288) 
(1600,150)+-(400,400)
(1388,140)+-(463,463)
(1355,130)+-(174,174)
(1857,120)+-(310,310)
(769,110)+-(73,73)
(296,100)+-(111,88) [blue]
(930,90)+-(240,240) 
(249,80)+-(51,51)
(331,70) -=(107,0) +=(161,0)
(1110,60)+-(225,225)
(750,50)+-(82,82)
(1130,40) -=(156,0) +=(156,0)
(328,30) -=(42,0) +=(38,0)
};
\draw[dashed,black] (370,-18.4) -- (370,215);
\node[rotate=90]  at (320,160) {\small{CMB}};
\node[rotate=90,text=gray]  at (640,99) {\footnotesize{$3\sigma$ upper bound}};
\end{axis}
\end{tikzpicture}

\vspace{-1mm}

\begin{tikzpicture}
    \begin{axis}[
        xmin=0,
        xmax=2280, 
        ymin=0, ymax=30, 
        line width=0.8pt,
        xlabel={$\lvert \boldsymbol{v}\rvert$ (km/s)},
        ytick={20,10},
        yticklabels={
        \scriptsize{$\;\;\;\;\;\;\;\;\;\;\;\;\,\,$[eBOSS]$\,\;\;\;\;\;$}, 
        \scriptsize{$\;\;\;\;\;\;\;\;\;\;\;\;\,\,$[BOSS]$\,\;\;\;\;\;$}}, 
        scale only axis=true,
        width=0.8\textwidth,
        height=0.075\textwidth,
        tick style={color=black},
        axis on top=true
    ]
    \fill[gray!20] (axis cs:0,-5.4) rectangle (axis cs:700,185.5);
    
    \addplot+[black, only marks, every mark/.append style={solid, fill=black,mark size= 1.5pt}, mark=square*,error bars/.cd, x dir=both,x explicit]
 coordinates { 
(196,20) -=(79,0) +=(92,0)
(184,10) -=(80,0) +=(101,0)
};
\draw[dashed,black] (370,-21.6) -- (370,205);
\node[rotate=0]  at (2000,19.8) {\small{This work}};
\end{axis}
\end{tikzpicture}
\caption{\textbf{Velocity estimates from dipoles in chronological order with $1\sigma$ error bars.}
\textit{Top:} previous results from x-ray, radio galaxies (RG), quasars (QSO) and SNe Ia, with corresponding references.
\textit{Bottom:} results of this work, combining several types of objects in the range $0.4<z<2.2$.
\textit{Dashed line:} the velocity obtained by the CMB dipole, considering the peculiar velocity hypothesis.
\textit{Gray region and 2021 CMB result:} the velocity measured by ref.~\citealp{Ferreira:2020aqa} using CMB non-diagonal correlations and its $3\sigma$ upper bound. 
}
\label{fig:dipoles_diffsources}
\end{figure}

Cosmology's standard model, the $\Lambda$ cold dark matter ($\Lambda$CDM) model, is founded on the cosmological principle, which asserts that the universe is statistically homogeneous and isotropic on scales larger than approximately 100 Mpc~\cite{Scrimgeour:2012wt,Ntelis:2017nrj}.
Consequently, the primary source of the dipole observed in distant objects should be attributed to our peculiar velocity $\boldsymbol{v}$ relative to the CMB rest frame.
This hypothesis underlies the Ellis-Baldwin test~\cite{1984MNRAS.206..377E}, which posits that velocity estimates derived from different sources, such as x-ray~\cite{Plionis:1998xj}, radio galaxies (RG)~\cite{Blake:2002gx,singaldip,Rubart:2013tx,Colin:2017juj,Bengaly:2017slg,Murray:2021frz,Darling:2022jxt,Secrest:2022uvx,Wagenveld:2023kvi}, quasars (QSO)~\cite{Secrest:2020has,Secrest:2022uvx}, supernovae Ia (SNe)~\cite{Horstmann:2021jjg,Sorrenti:2022zat}, and even gravitational waves~\cite{Mastrogiovanni:2022nya,Tasinato:2023zcg}, should agree.
However, over the past 25 years, several studies utilizing different RG and QSO catalogs have consistently found that, though the peculiar velocity is generally consistent in direction with the CMB dipole, its amplitude is in tension, with a value of $\lvert \boldsymbol{v}\rvert\sim1000$ km/s, as reviewed in Fig.~\ref{fig:dipoles_diffsources}. This quarter-century-long puzzle poses a significant challenge to the $\Lambda$CDM model, with the tension regarding the CMB dipole now reaching a $5\sigma$ level~\cite{Secrest:2022uvx,Wagenveld:2023kvi}, suggesting that the CMB rest frame is not shared by all matter in the observable universe. It warrants serious consideration, as any deviations from large-scale homogeneity and isotropy, if confirmed, would fundamentally alter our understanding of the cosmos and the primordial physics that set the universe's initial conditions.

The possibility exists that part of the CMB dipole might originate from intrinsic cosmological phenomena, such as primordial fluctuations on the last-scattering surface~\cite{Roldan:2016ayx}, the presence of a large local void~\cite{Paczynski1990}, ``tilted universe'' scenarios~\cite{Turner_1992,Domenech:2022mvt}, and specific inflationary models~\cite{Langlois:1996ms}. This suggests that the CMB dipole might not solely result from our peculiar velocity relative to the CMB rest frame, but could also include an intrinsic cosmological component, which could account for discrepancies between velocities estimated from the CMB and those derived from other astronomical sources.
However, by following the footprints of aberration and Doppler in harmonic space~\cite{Ferreira:2021omv}, ref.~\citealp{Ferreira:2020aqa} put the first  constraint on the intrinsic CMB dipole 
and the findings are in agreement with the kinematic interpretation supported by the $\Lambda$CDM model. Before ref.~\citealp{Ferreira:2020aqa}, all the previous results from Fig.~\ref{fig:dipoles_diffsources} could accommodate the observed CMB dipole if one allows a fine-tuned scenario with a non-negligible intrinsic dipole in the opposite direction of our peculiar velocity. Such a situation implies a reduced total dipole, 
which would be wrongly interpreted as a smaller velocity.
This is not more a possibility, as can be seen in Fig.~\ref{fig:dipoles_diffsources} where we show the $3\sigma$ upper limit, updated according to ref.~\citealp{Murray:2021frz}, on our peculiar velocity as determined by ref.~\citealp{Ferreira:2020aqa} using Planck's temperature and polarization data and assuming a non-zero intrinsic dipole component.

Recent measurements using SNe data~\cite{Horstmann:2021jjg,Sorrenti:2022zat} reported velocities below those expected by the CMB dipole, suggesting that systematic effects or an unexpectedly large clustering dipole could have an important role in this tension. Supporting this perspective, ref.~\citealp{Mittal:2023xub} showed that the number count dipole of the new Quaia catalog of QSOs from Gaia could align with the CMB dipole, depending on the galactic cut applied. Additionally, ref.~\citealp{Siewert:2020krp} reported a radio dipole amplitude exceeding CMB predictions, with the unexpected characteristic of amplitude increasing inversely with frequency. For x-ray, we will need to wait for the next results using the eROSITA satellite data~\cite{merloni2012erosita} to see if they will converge to a smaller~$\lvert \boldsymbol{v}\rvert$.
Moreover, ref.~\citealp{Rubart:2013tx} draws attention to the fact that to convert directly the measured dipoles to the frame velocity, i.e., to considering it purely kinetic, is a naive approximation that does not consider biases due mainly to shot noise and masking.

Usually, only number counts of objects are used to measure the dipole modulation. However, ref.~\citealp{Nadolny:2021hti} demonstrated that Doppler and aberration effects also introduce measurable modulations in the flux, redshift, and angular size of objects. Our goal is to employ the extensive and precise spectroscopic redshift measurements from quasars and galaxies provided by the Sloan Digital Sky Survey (SDSS) to estimate the dipole tomographically across redshift bins and to verify the validity of the cosmological principle, that is, whether the matter frame of galaxies and quasars coincides with the CMB rest frame.

\section{The redshift dipole}

A Lorentz boost changes the redshift $z$ of a source in the direction $\nhat$ to the observed redshift~$z^\prime$ in the observed direction $\nhatp$ by
\begin{equation}
    1+z^\prime (\nhatp) = [1+z(\nhat)] \delta(\boldsymbol{\beta},\nhatp) \;,
\label{zboosted}
\end{equation}
where $\bm{\beta}= \boldsymbol{v}/c$, $\boldsymbol{v}$ is our velocity, $c$ is the speed of light, and $\delta(\boldsymbol{\beta},\nhatp)$ is the Doppler factor defined by
\begin{equation}\label{eq:doppler_factor1}
     \delta(\boldsymbol{\beta},\nhatp) = \frac{\sqrt{1-\beta^2}}{(1+\boldsymbol{\beta} \cdot \nhatp)} \;.
\end{equation}
The change in observed direction due to aberration is given by
\begin{equation}
    \hat{\boldsymbol{n}}^{\prime}=\frac{\hat{\boldsymbol{n}} \cdot \hat{\boldsymbol{\beta}}+\beta}{1+\hat{\boldsymbol{n}} \cdot \boldsymbol{\beta}} \, \hat{\boldsymbol{\beta}}+\frac{[\hat{\boldsymbol{n}}-(\hat{\boldsymbol{n}} \cdot \hat{\boldsymbol{\beta}}) \hat{\boldsymbol{\beta}}]}{\gamma(1+\boldsymbol{\beta} \cdot \hat{\boldsymbol{n}})} \;,
\end{equation} such that, for an intrinsically isotropic redshift distribution in the sky, by expanding Eq.~\eqref{zboosted} in multipoles, the redshift dipole is~\cite{Nadolny:2021hti} 
\begin{equation}
    \bm{\Delta_1} = -\bm{\beta} + \mathcal{O}(\beta^2) \; . \label{eq:delta_1}
\end{equation}
This implies that objects in the direction of the boost experience a reduction in redshift, whereas those in the opposite direction see an increase. Considering the peculiar velocity hypothesis, the expected $\Delta_1$ is of order $\sim10^{-3}$, corresponding to a maximum effect of $\sim0.1\%$ in the redshift. 

An observer at rest relative to the Hubble flow can still observe a dipole anisotropy arising from the universe's large-scale structure. This effect, known as the clustering dipole, becomes increasingly significant at lower redshifts ($z \ll 1$), as fluctuations in density and velocity are more pronounced at smaller scales.
This intrinsic contribution must be added to Eq.~\eqref{eq:delta_1}, leading to
\begin{equation}
    \bm{\Delta_1} = -\bm{\beta} + \bm{\Delta_{1,\rm{int}}} + \mathcal{O}(\beta^2) \; .
\end{equation}
The clustering dipole's main contribution comes from peculiar motions, and can be estimated by computing the bulk flow velocity, which is the expected root mean square velocity of all sources in a survey (see equation D.2 from \cite{Nadolny:2021hti} for the exact definition).
We found that the clustering dipole is negligible for the dataset we consider, with a magnitude less than 2\% of the expected $\beta$~\cite{Nadolny:2021hti}. Consequently, we will adopt $\bm{\Delta_1} = -\bm{\beta}$, thereby disregarding terms of the order $\mathcal{O}(\beta^2)$. In our approximation, where the only source of the dipole is the velocity, this implies a unique definition of $\boldsymbol{\beta}$ as $\boldsymbol{\beta} = -\boldsymbol{\Delta_1}$, ensuring that the dipole is null in the rest frame.

\section{Data}\label{sec:data}

\begin{figure}
\centering 
\includegraphics[trim={7 0 0 0}, clip, width=.7 \columnwidth]{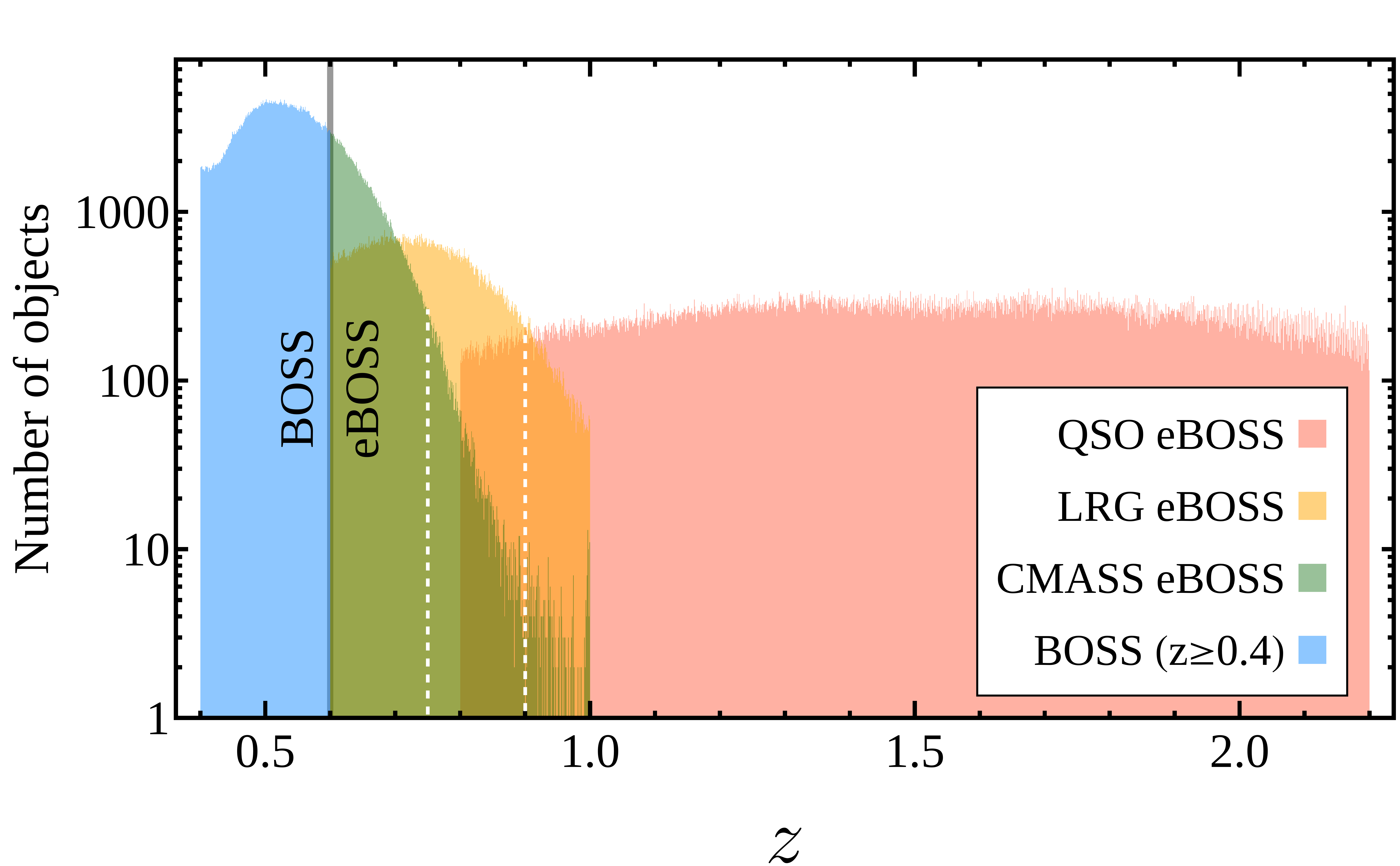}
\caption{\textbf{Redshift distribution of SDSS data used.}
The bin size adopted in the plot is $\Delta z = 0.001$. Dashed lines indicate data constraints. From left to right, the first dashed line indicates the upper bound for CMASS eBOSS data, $z<0.75$, implemented to prevent estimator misbehavior with sparsely populated bins. Similarly, for LRG data, we adopt $z<0.9$, denoted by the second dashed line.
}
\label{fig:z_hist}
\end{figure}

\begin{figure}
\centering 
\includegraphics[trim={0 10 0 -29.5mm}, clip, width=0.49\columnwidth]{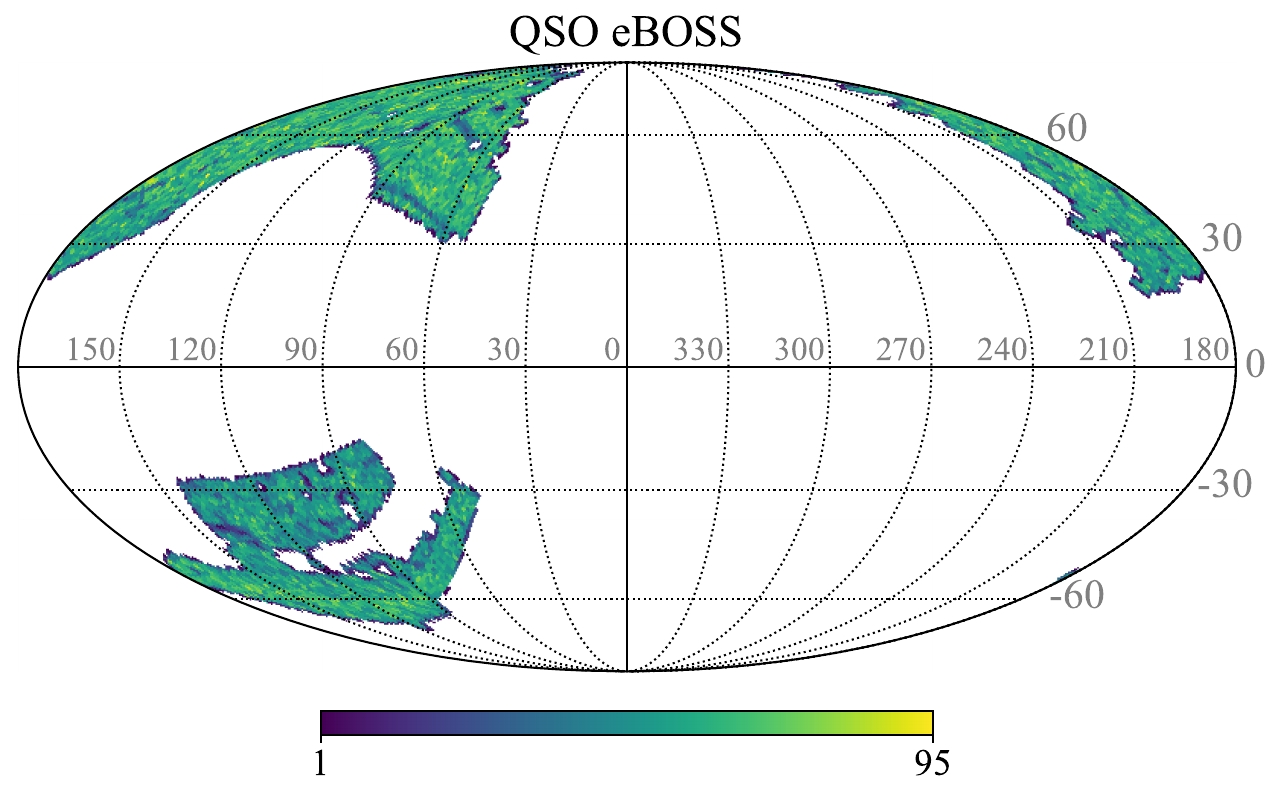}
\includegraphics[trim={0 10 0 0mm}, clip, width=0.49\columnwidth]{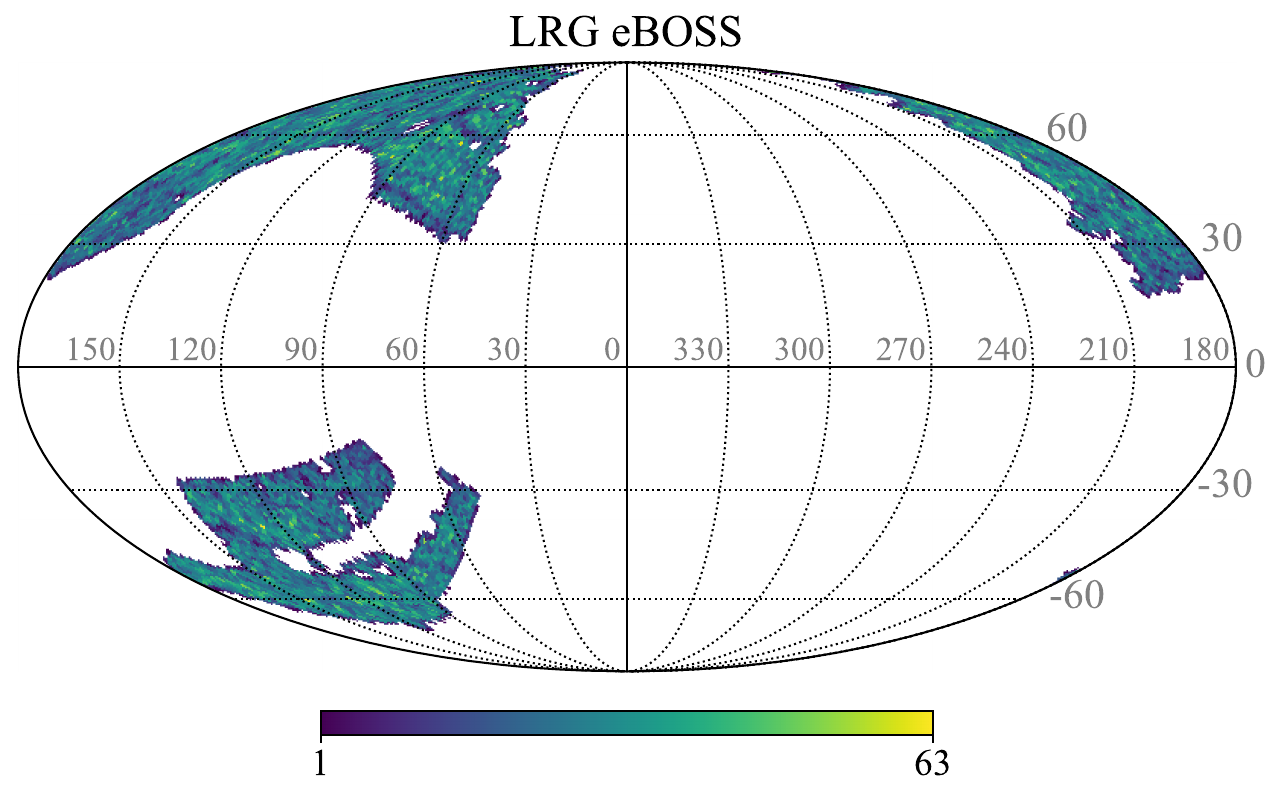}
\includegraphics[trim={0 10 0 -1mm}, clip, width=0.49\columnwidth]{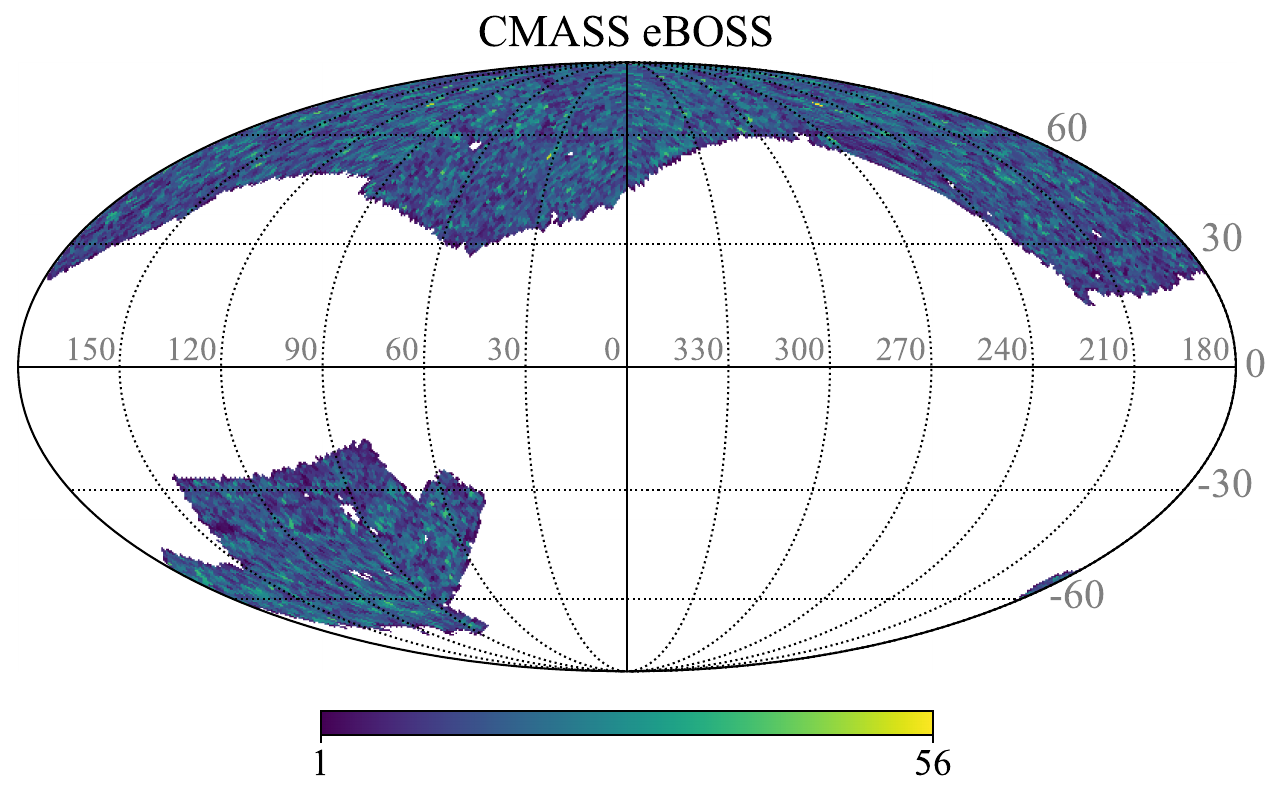}
\includegraphics[trim={0 10 0 -2mm}, clip, width=0.49\columnwidth]{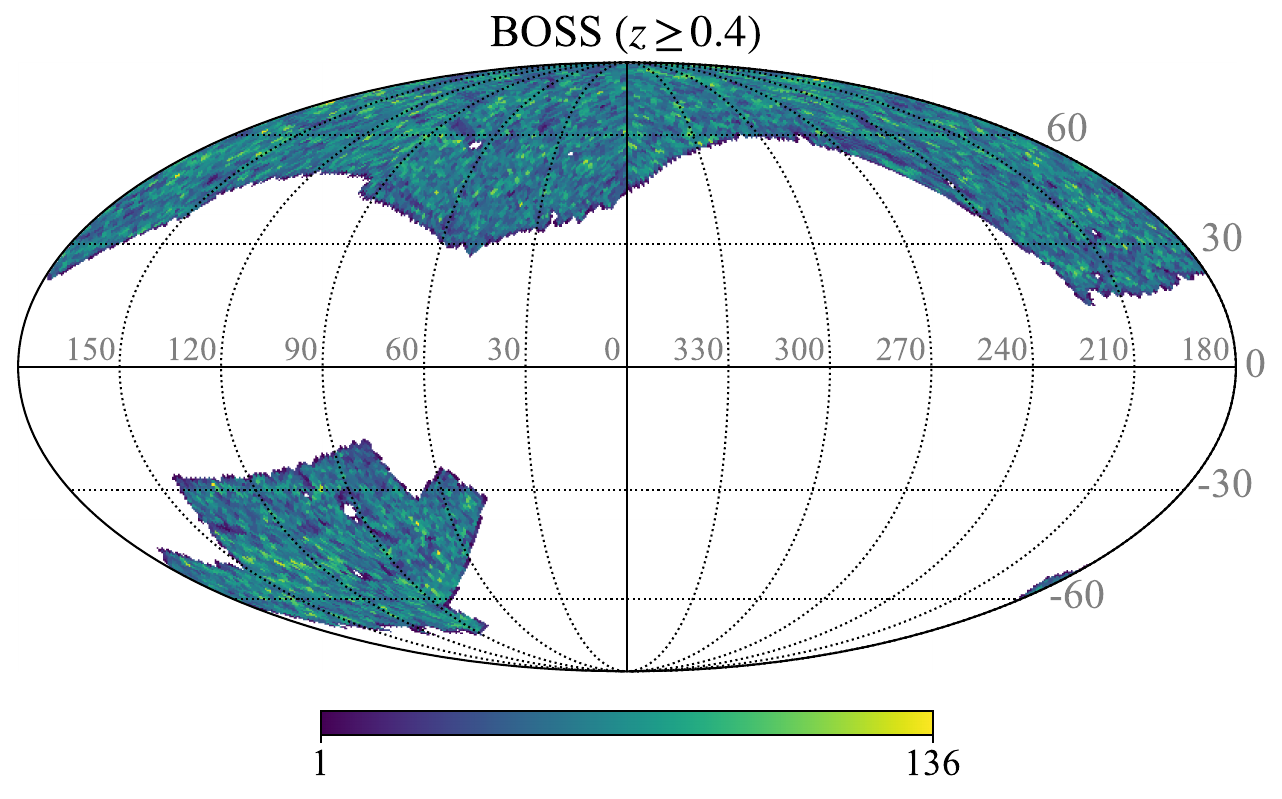}
\caption{\textbf{Tracer distribution footprints.} Number count maps for each tracer, presented in Mollweide projection using the \texttt{HEALPix}  format with $N_{\rm side}=64$.}
\label{fig:footprints}
\end{figure}

We used tracers from 4 different catalogs from eBOSS~\cite{eBOSS:2020mzp} and BOSS~\cite{Reid:2015gra} SDSS programs:
i) the QSO DR16 eBOSS catalog in the range ${0.8\!<\!z\!<\!2.2}$, encompassing a fraction of sky $f_{\rm sky}=0.14$ with $343708$ objects;
ii) the Luminous Red Galaxies DR16 eBOSS catalog (LRG eBOSS) in the range ${0.6\!<\!z\!<\!0.9}$, covering a $f_{\rm sky}=0.14$ with $163249$ objects;
iii) the massive galaxies DR16 eBOSS catalog (CMASS eBOSS) in the range $0.6\!<\!z\!<\!0.75$, embracing $f_{\rm sky}=0.25$ and $193298$ objects;
iv) the galaxies DR12 BOSS catalog in the range $0.4\!<\!z\!<\!0.6$ (mostly CMASS galaxies), with $f_{\rm sky}=0.26$ and $686370$ objects.
See Fig.~\ref{fig:z_hist} for the catalogs' redshift histograms and Fig.~\ref{fig:footprints} for the footprints. All these catalogs are divided into North Galactic Cap (NGC) and South Galactic Cap (SGC), and provide the weights to correct for imaging systematics, fiber collisions and redshift failures.

To infer the errors and biases of the estimator for eBOSS data, we used the 1000 EZmock realistic samples~\cite{Chuang:2014vfa,Zhao:2020bib}, which include effects of clustering, survey geometry, redshift evolution, sample selection biases, and all systematic effects. These samples were modified to incorporate the Doppler effect by applying Eq.~\eqref{zboosted}. For the BOSS catalogs, we applied the same approach using 1000 MultiDark-Patchy mocks~\cite{Kitaura:2015uqa}. Both mock pipelines have been employed by the SDSS collaboration for comprehensive analysis of baryon acoustic oscillations and redshift space distortions~\cite{BOSS:2016wmc,eBOSS:2020gbb,eBOSS:2020hur}.

The Emission Line Galaxy (ELG) eBOSS catalog was not considered as it covers merely $\sim$3\% of the sky. Given our aim to assess the largest scale modulation, the extent of sky coverage is crucial; hence, incorporating the ELG catalog yields negligible enhancement to our analysis.
BOSS objects with $z<0.4$ (mostly LOWZ galaxies) were not considered because it was not possible to minimize the estimator. We will come back to this issue in Section~\ref{results}.
Furthermore, CMASS eBOSS objects with $z>0.75$ were excluded, leading to a 4.6\% reduction in the original catalog, and LRG eBOSS objects with $z>0.9$ were also omitted, accounting for a 6.6\% reduction. These minor exclusions were necessary to prevent estimator misbehavior, as the estimator's minimization becomes unstable in sparsely populated redshift bins.

\section{Measuring the dipolar modulation}

\subsection{Estimator}

Achieving an unbiased estimation of the dipole and a reliable assessment of its uncertainty presents a significant challenge, especially when targeting a subtle signal of $\Delta_1\sim10^{-3}$. To tackle this issue, we introduce a novel dipole estimator, implement a comprehensive bias-removal pipeline, and quantify uncertainty using state-of-the-art SDSS mock catalogs. We will now discuss the estimator and the three key elements of our measurement technique.

We employ a least squares estimator to measure the dipole $\bm{\Delta}_{1}$ for each bin, based on the peculiar velocity hypothesis $\bm{\Delta}_{1} = -\bm{\beta}$.
This calculation involves comparing the observed redshift distribution to the expected one for a theoretically isotropic $z$-field, boosted as per Eq.~\eqref{zboosted}:
\begin{equation}
\chi^2_{\rm bin} (\bm{\Delta}_{1}) = \frac{ \sum_{i}^{N} w_i \left[1+ z_{i} - \Delta^{dD}_{0,{ \rm bin}} \delta(-\bm{\Delta}_{1},\bm{\hat{n}}^{\prime}_{i}) \right]^2}{\sum_{i}^{N} w_i} ,
\label{eq:estimator}
\end{equation}
where $z_i$ is the observed redshift of object $i$ in a specific bin, $\bm{\hat{n}}^{\prime}_{i}$ is the observed redshift direction, $N$ denotes the total number of objects, and $w_i$ are the weights for each catalog (eBOSS and BOSS) defined as:
\begin{align}
    w_{i,\mathrm{eBOSS}} &= w_{i,\mathrm{sys}} \cdot w_{i,\mathrm{noz}} \cdot w_{i,\mathrm{fc}}  \,, \\
    w_{i,\mathrm{BOSS}} &= w_{i,\mathrm{sys}} \cdot (w_{i,\mathrm{noz}}+w_{i,\mathrm{fc}}-1) \,,
\end{align}
accounting for redshift failures ($w_{i,\mathrm{noz}}$), fiber collisions ($w_{i,\mathrm{fc}}$), and imaging systematics ($w_{i,\mathrm{sys}}$).

\begin{figure}
\centering 
\includegraphics[trim={0 0 0 0}, clip, width=\columnwidth]{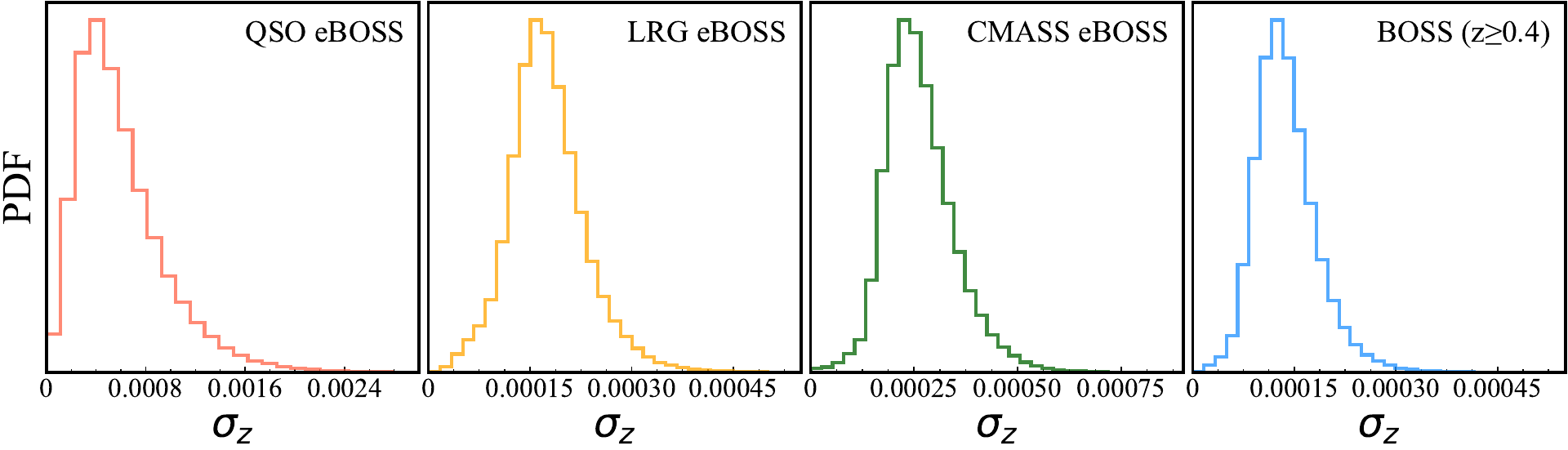}
\caption{\textbf{Redshift error distribution for each sample.}
The median error from left to right is 0.00052, 0.00017, 0.00025, 0.00013.}
\label{fig:z_error}
\end{figure}

The spectroscopic redshift measurement errors are approximately $2 \times 10^{-4}$ for LRG eBOSS, CMASS eBOSS, and BOSS $(z\geq0.4)$, and around $5 \times 10^{-4}$ for QSO eBOSS objects~\cite{2012AJ....144..144B,Reid:2015gra,Paris:2016xdm}, as shown in Fig.~\ref{fig:z_error}. The 90th percentiles of the error distributions for QSO eBOSS, LRG eBOSS, CMASS eBOSS, and BOSS $(z\geq0.4)$ are 0.0011, 0.00025, 0.00037, and 0.00020, respectively. Consequently, we conservatively choose redshift bin widths that are approximately three times these values, adopting $\Delta z = 0.001$ for LRG eBOSS, CMASS eBOSS, and BOSS $(z\geq0.4)$, and $\Delta z = 0.003$ for QSO eBOSS. As verified in Appendix~\ref{revo}, the error evolution for the QSO eBOSS sample across different bins remains consistent with the chosen bin size throughout the entire redshift range.

\begin{figure}[t]
\centering 
\includegraphics[trim={0 0 0 0}, clip, width=.5\columnwidth]{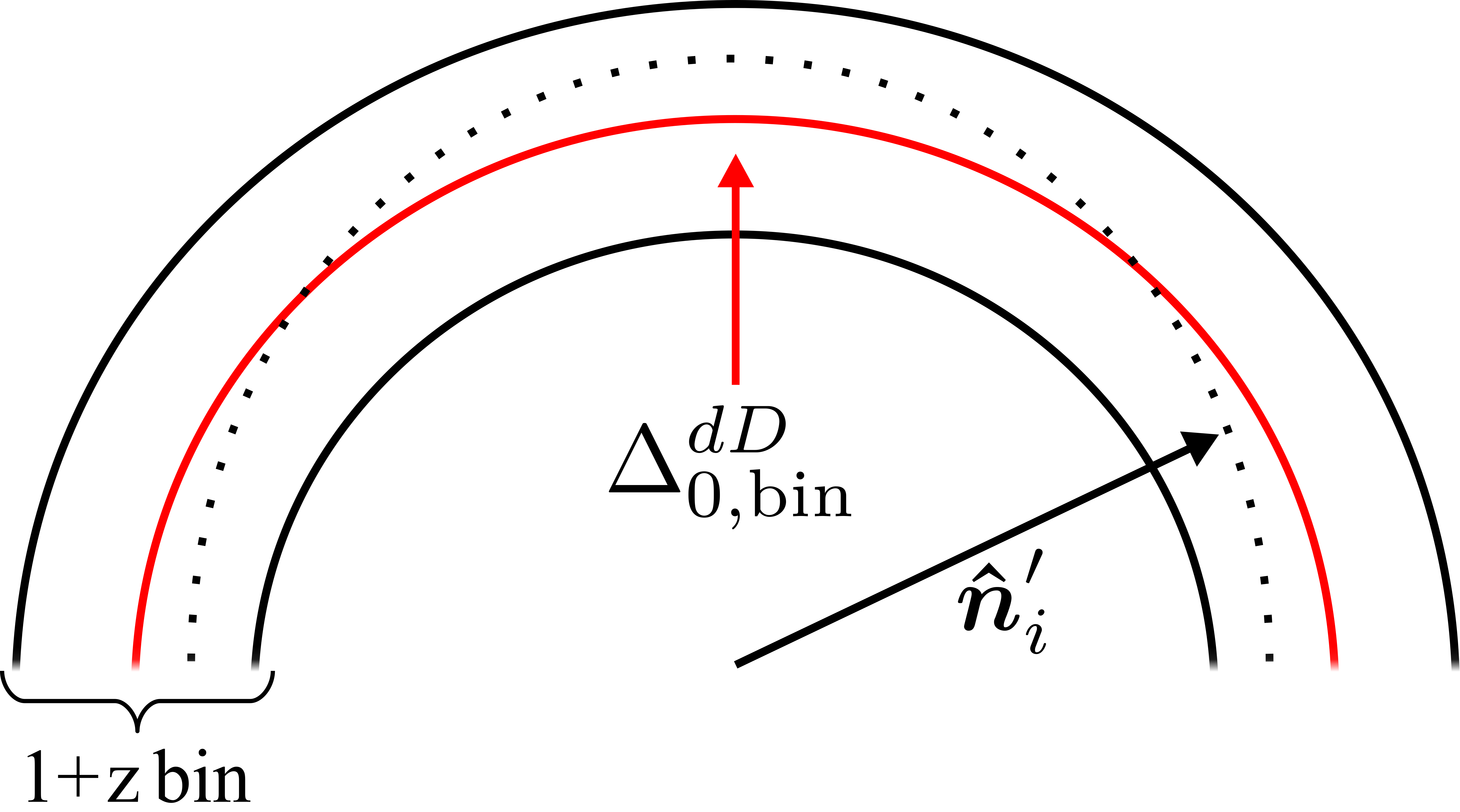}
\caption{\textbf{Estimator diagram.} Diagram of the $1+z$ spherical shell bin, its monopole and data.The dotted curve represents the observed 1+z for several objects (dots). The estimator minimizes the difference between the observed $1+z_i$ (black points) and the monopole $\Delta^{dD}_{0,{ \rm bin}}$ (the average of $1+z_i$ in the rest frame, described by the red line) times a dipole modulation $\delta(-\bm{\Delta}_{1},\bm{\hat{n}}^{\prime}_{i})$. The direction vector $\bm{\hat{n}}^{\prime}_{i}$ originates from the observer at the center of the shell. 
}
\label{fig:estimator_diagram}
\end{figure}

The first key point regards the estimation of the monopole itself.
In a full-sky survey, the monopole and de-Dopplered monopole are expected to be identical because Doppler modulation symmetrically increases the redshifts in one direction while decreasing them in the opposite direction. However, this antipodal symmetry is disrupted in any sky-limited survey. To avoid introducing bias, it is necessary to use the ``de-Dopplered'' monopole $\Delta^{dD}_{0,{ \rm bin}}$ of the respective bin, which is the average $1+z$ with the Doppler modulation removed,
\begin{equation}
    \Delta^{dD}_{0,{ \rm bin}} = \sum_{i}^{N} w_i (1+z_{i}) \delta(-\bm{\Delta}_{1},\bm{\hat{n}}^{\prime}_{i})^{-1} \bigg/ \sum_{i}^{N} w_i \,,
\end{equation}
that is, the original rest-frame monopole according to the tested value of $\bm{\Delta}_{1}$.
Summarizing, our estimator considers objects individually, with their respective systematic weights, comparing the observed and expected boosted redshift for its coordinates, given the de-Dopplered monopole of the very narrow redshift bin and the velocity being tested.
See Fig.~\ref{fig:estimator_diagram} for an illustration of the estimation procedure. 

\begin{figure}[t]
\centering 
\includegraphics[trim={0 0 0 0}, clip, width=.63\columnwidth]{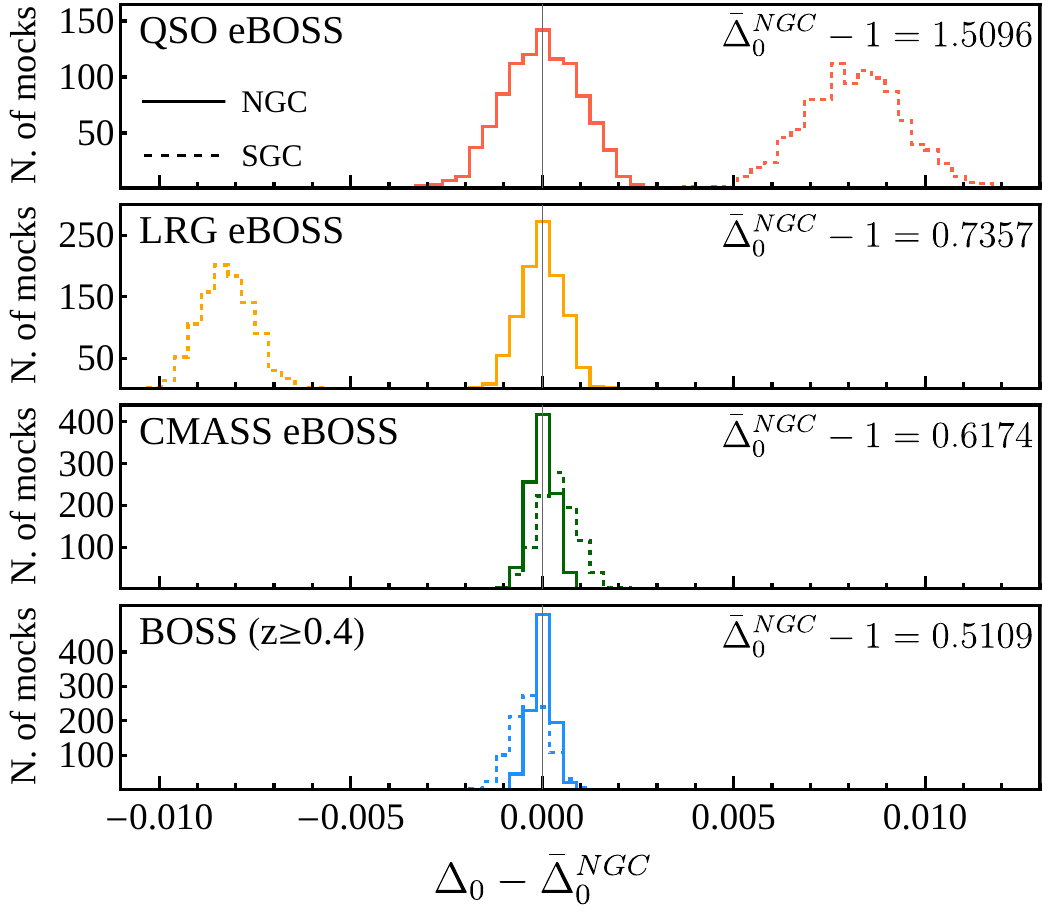}
\caption{\textbf{Distribution of monopoles ($\Delta_0$, average redshift) for the North (NGC) and South (SGC) Galactic Caps mocks.} $\bar{\Delta}^{\rm NGC}_0$ represents the average NGC monopole, i.e., average $1+z$. It is clear that the hemispheres have different depths in the QSO eBOSS and LRG eBOSS samples. Even for CMASS eBOSS and BOSS $(z\geq0.4)$ the difference is not insignificant for our estimator, as it provides a systematic shift of $\sim3.5\times10^{-4}$, while the expected modulation is $\sim10^{-3}$.
}
\label{fig:monopoles}
\end{figure}

The second key point concerns employing a separate analysis for the two galactic caps. 
Despite both caps being designed with equivalent equipment, analytical techniques, and observational strategies, a detailed examination of mocks and observational data reveal distinct depth variations between the hemispheres. As a result, the redshift monopoles $\Delta_0$ (average $1+z$ considering the systematic weights) differ, as demonstrated in Fig.~\ref{fig:monopoles}. These variations are comparable to, or exceed, the magnitude of the expected dipole $\Delta_1$ ($\sim10^{-3}$). Given that the dipole is evaluated as a modulation around the monopole, acknowledging these discrepancies is essential to prevent bias in the combined estimation from the NGC and SGC datasets. 
Indeed, measuring the dipole, with and without hemisphere division, shows a typical deviation around $600$ km/s for individual catalogs. When analyzing the two galactic caps together, the Doppler modulation is adjusted around the combined monopole of both hemispheres, with differences of the same order or bigger than the $\Delta_1 c$ value of $\sim$370 km/s, expected within the $\Lambda$CDM model. Accounting for this effect is crucial for an accurate measurement. That is why we analyze objects in the Northern and Southern Galactic Caps separately. 
Therefore, to find the complete sample's best-fit dipole $\bm{\Delta}_1$, we minimize across all bins according to:
\begin{equation}
    \chi^2 (\bm{\Delta}_{\rm 1}) = \sum_{ \rm bin}  \chi^2_{\rm bin, NGC} (\bm{\Delta}_{1})+ \sum_{ \rm bin}  \chi^2_{\rm bin,  SGC} (\bm{\Delta}_{1})\,.
\end{equation} The final result is obtained by minimizing across all bins and hemispheres simultaneously. 

\subsection{Bias estimation and removal}
\label{bias}

The third key point of our analysis pipeline addresses estimating and removing potential biases in the dipole measurement. Systematic survey errors, including inconsistencies in observational strategies, footprint inhomogeneities, and variations in sky and telescope conditions, can introduce significant biases. To assess and correct for these biases, we utilized mock catalogs derived from eBOSS's EZmock sample and BOSS's MultiDark-Patchy mocks. Notably, the realistic version of EZmock incorporates observational systematics, a feature absent in the MultiDark-Patchy mocks, potentially affecting the findings related to the BOSS catalog.

For the BOSS mocks, which do not account for systematic effects as EZmocks do, we defined the weights for bias estimation as follows:
\begin{equation}
   w^{\rm{mocks}}_{i,\mathrm{BOSS}} = w_{i,\mathrm{veto}} \cdot w_{i,\mathrm{fc}},
\end{equation}
where $w_{i,\mathrm{veto}}$ serves as a flag, set to 1 for objects within the ``veto mask'' and 0 for those outside. The veto mask is designed to exclude sky sections not meeting specific quality standards, thereby minimizing areas heavily influenced by systematics. These standards may include regions with insufficient observational coverage, areas affected by contamination, or other conditions that could compromise data quality.

\begin{figure}
\centering 
\includegraphics[trim={0 0 0 0}, clip, width=\columnwidth]{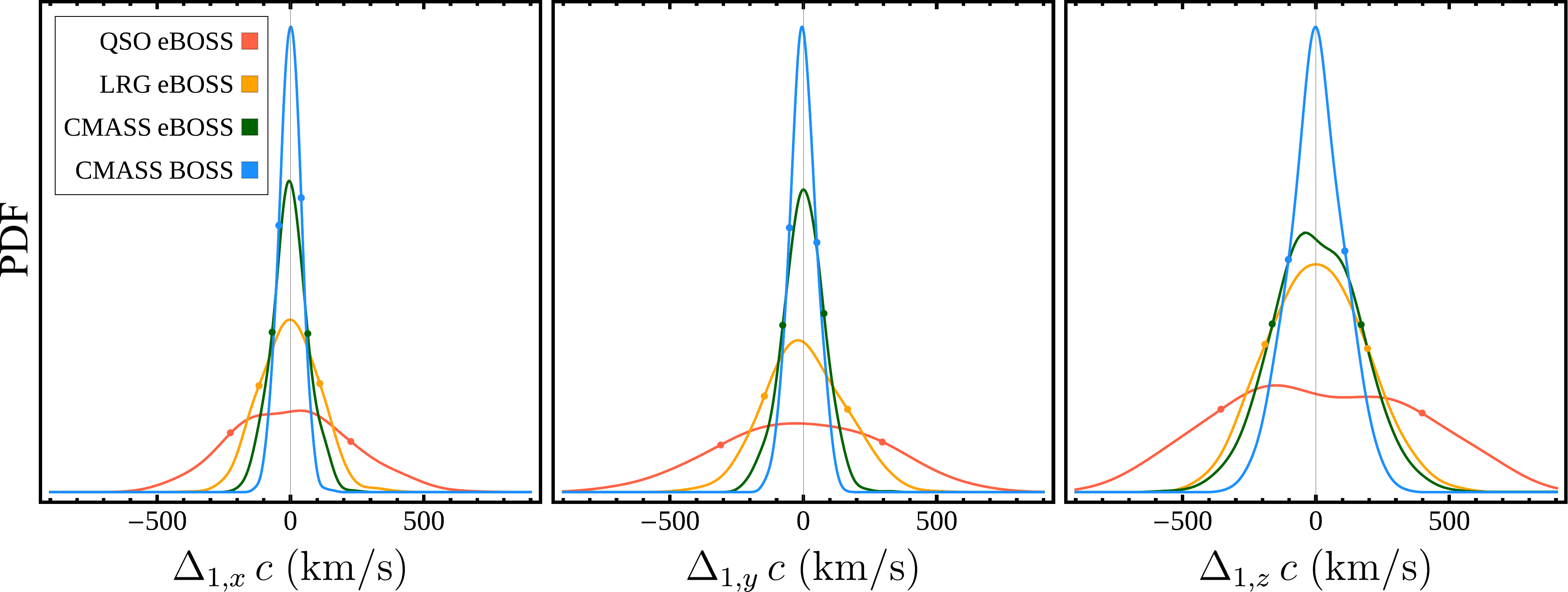}
\caption{\textbf{Probability distributions of de-biased results without boost}. The points mark the $1\sigma$ confidence intervals.}
\label{fig:null_case}
\end{figure}

To construct the bias map for both the eBOSS and BOSS surveys, we first Doppler boosted 100 mocks, randomly selected from the original pool of 1000 for each dataset, across 12 distinct directions identified by the pixel centers in the \texttt{HEALPix} $N_{\rm side}=1$ scheme\footnote{\texttt{HEALPix}~\cite{2005ApJ...622..759G,Zonca2019}, or Hierarchical Equal Area isoLatitude Pixelization, is an efficient system widely used in astrophysics and cosmology for pixelating the celestial sphere, enhancing the analysis and visualization of astronomical datasets.}. These mocks were simulated at a velocity of 370 km/s, by multiplying the $1+z$ value of each object by the expected Doppler factor as defined by equation \eqref{eq:doppler_factor1}, consistent with the kinematic interpretation of the CMB dipole. Our tests confirm that the bias map remains reliable across various dipole magnitudes. Additionally, we validated the accuracy of our estimator using the 4000 original mocks (1000 per catalog) without Doppler effects, successfully achieving expected results with bias correction, as depicted in Fig.~\ref{fig:null_case}.

\begin{figure}[t!]
\centering 
\includegraphics[trim={0 0 0 0}, clip, width=\columnwidth]{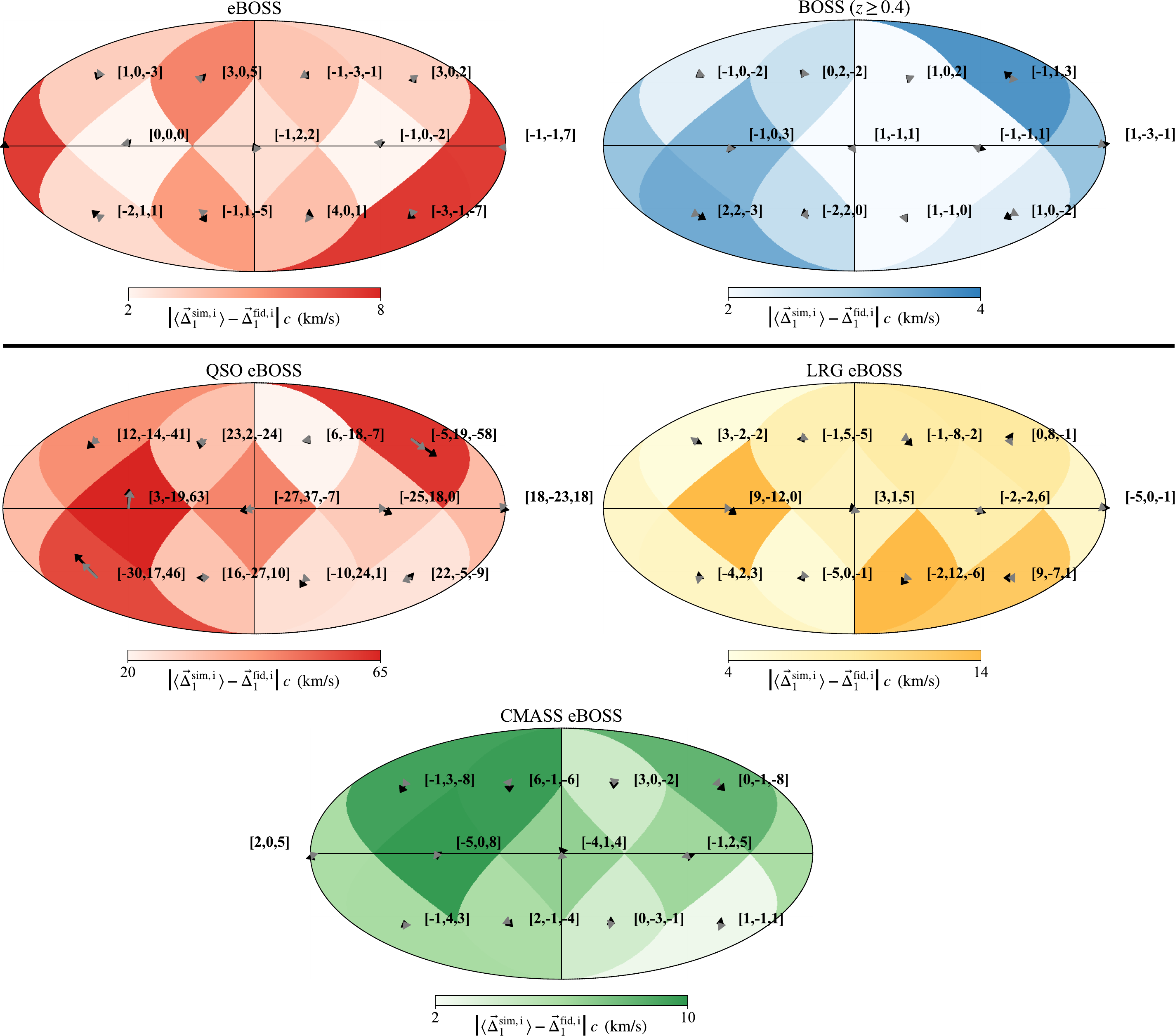}
\caption{\textbf{Bias maps in Mollweide projection.} For each direction $i$, arrows originate from the fiducial expected values $\bm{\Delta}^{\rm{fid,i}}_1$, defined at the center of the pixels in the \texttt{HEALPix} $N_{\rm side}=1$ scheme, and point towards the direction of the average estimated vector from simulations $\bm{\Delta}^{\rm{sim,i}}_1$. Grey arrows depict the de-biased results, whereas black arrows represent the biased outcomes. The numbers indicate the difference vector between the de-biased results and the fiducials, with colors denoting the absolute value of this vector, all expressed in km/s.}
\label{fig:bias_maps}
\end{figure}

For the bias vector calculation, we estimated the dipole for each of the 100 simulations per direction (1200 simulations per tracer). We computed the average difference vectors by subtracting the fiducial dipole vector $\bm{\Delta}^{\rm{fid,i}}_1$ for direction $i$ from the simulations' average dipole vector $\langle\bm{\Delta}^{\rm{sim,i}}_1\rangle$. These vectors, expressed through their Cartesian components, define the pixel values of the three bias maps for each tracer.
It is noteworthy that the detected bias is relatively minor.
We performed bi-linear interpolation of these bias maps using the \texttt{get\_interp\_val} function from the \texttt{healpy} Python module~\cite{Zonca2019}. This interpolation allows us to de-bias the measured dipole, ensuring precise analysis and interpretation. Fig.~\ref{fig:bias_maps} illustrates this process for both combined and individual cases, displaying the average biased and de-biased dipole vectors against the benchmark dipoles from the simulations. The corrected bias for each Cartesian component aligns with the benchmark values, confirmed by comparing the mean difference vectors against the standard errors derived from 100 mocks per direction.

\subsection{Correlations}

When observed within the same redshift range, different tracers are expected to exhibit a similar clustering dipole, $\bm{\Delta_{1,\rm{int}}}$, which could lead to correlations between measurements not accounted for by our estimator. Despite our theoretical assessment indicating that the intrinsic dipole's influence is minor (less than 2\% of the anticipated dipole), we employed the eBOSS mock datasets for a practical evaluation.

\begin{figure}
\centering 
\includegraphics[trim={0 0 0 0}, clip, width=0.9\columnwidth]{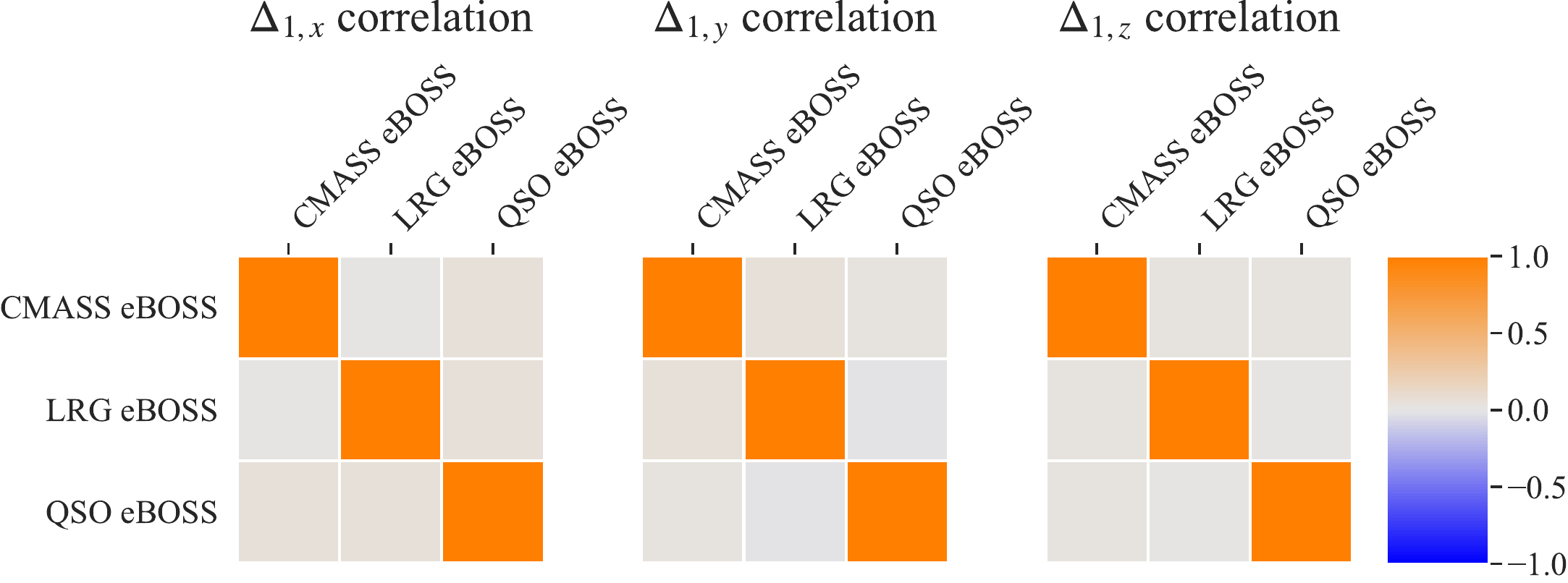}
\caption{\textbf{Pearson correlation matrix between $\Delta_1$ measurements of different tracers.}
The correlations were obtained using the eBOSS mocks.}
\label{fig:correlations}
\end{figure}

Using these datasets, we assessed whether different tracers within the same redshift range demonstrate a comparable clustering dipole, attributable to sampling the same underlying dark matter density and velocity field.  This effect potentially induces correlations between measurements. EZmock individual tracers' mocks, sharing identical ID numbers, originate from the same dark matter realization, ensuring they exhibit a similar clustering dipole within the same redshift range. We evaluated the dipole for each simulation
and calculated the correlation between different catalogs, finding that, as illustrated in Fig.~\ref{fig:correlations}, the correlation is insignificant.
This implies that the primary source of uncertainty is attributed to the finite tracers' number density. Our empirical results reaffirm the theoretical prediction, establishing that the clustering dipole's effect is indeed negligible.

\subsection{Error estimation}

To accurately estimate the uncertainty associated with the dipole estimation, we applied a Doppler boost to the 1000 BOSS and eBOSS mock catalogs based on the best-fit model derived from the real dataset (for a total of 4000 mocks). We then processed these boosted mocks through the complete pipeline, effectively sampling the measurement distribution to determine the confidence intervals.
Notably, the distributions obtained are consistent with the unbiased best-fit dipoles, serving as a further critical consistency check of the effectiveness of our bias removal technique across various dipole amplitudes, demonstrating that the bias correction is not significantly influenced by the absolute value of the dipole, including those different from the CMB-derived expectation used to create the bias maps.

Given our utilization of a finite set (1000) of mock catalogs for statistical analysis, we ensure the robustness of our findings by employing two complementary methods, both leading to similar conclusions: a non-parametric Kernel Density Estimation (KDE) with a Gaussian kernel and a parametric best-fit distribution analysis using the \texttt{Distfit} Python module~\cite{Taskesen2020}. This approach was consistent across error estimation and the determination of two-tailed significance regarding potential discrepancies with the CMB dipole. Our primary analysis leverages KDE.
In agreement with our frequentist methodology for assessing uncertainties using mock catalogs, central values and $1\sigma$ uncertainties are determined using quantiles: the 50th percentile for the central value, and the 16th and 84th percentiles for the lower and upper uncertainty bounds, respectively.

\subsection{Data combination}

To combine the results relative to the individual tracers and obtain the eBOSS results, we implemented the following steps:
i) We determined the probability distributions  $\{P^{\rm case}_x, P^{\rm case}_y, P^{\rm case}_z\}$ for each Cartesian component of the dipole vector $\{\Delta^{\rm case}_{1,x}, \Delta^{\rm case}_{1,y}, \Delta^{\rm case}_{1,z}\}$,  across all individual cases;
ii) We then multiplied the distributions of the Cartesian components together, followed by renormalization. For instance, prior to renormalization,
\begin{align} 
    P^{\text{\tiny{eBOSS}}}_x \!&=\! P^{\text{\tiny{QSO eBOSS}}}_x\!\times\! P^{\text{\tiny{LRG eBOSS}}}_x\!\times\!P^{\text{\tiny{CMASS eBOSS}}}_x \;\,,  \label{combo}
\end{align}
for combining the distributions of the $x$ component;
iii) Utilizing the combined distributions, we sampled 10000 vectors for the subsequent analysis.

The bias removal method of Section~\ref{bias} yields bias maps for individual tracers. To construct bias maps for the aggregate analysis of eBOSS, we adopt the strategy used for combining data, as depicted in Eq.~\eqref{combo}. Initially, we correct the 100 mock dipole measurements per direction using the bias maps specific to each tracer, then we aggregate these distributions and compare the composite average dipole to the fiducial one. The resultant difference vector forms the basis of the bias map for the combined analysis, which we subsequently interpolate to correct the measured dipoles derived from data combinations.


\section{Results and Discussion}
\label{results}

The analysis pipeline used in this study, including the evaluation of potential tension with the CMB dipole, was exclusively developed and validated using mock catalogs, ensuring that real data remained unexamined until after the methodology was established. Maintaining data blinding is critical to avoid reliance on \textit{a posteriori} statistics, which might inadvertently skew the findings.
In total, we processed 15000 mock catalogs, each subdivided into thousands of finely segmented redshift bins. The computational demand of this extensive study amounted to approximately 250000 CPU hours. 
For a visual representation of the data analysis pipeline, please refer to Appendix~\ref{pipe}.

\begin{figure}[t!]
\centering 
\includegraphics[trim={0 0 0 0.3mm}, clip, width=\columnwidth]{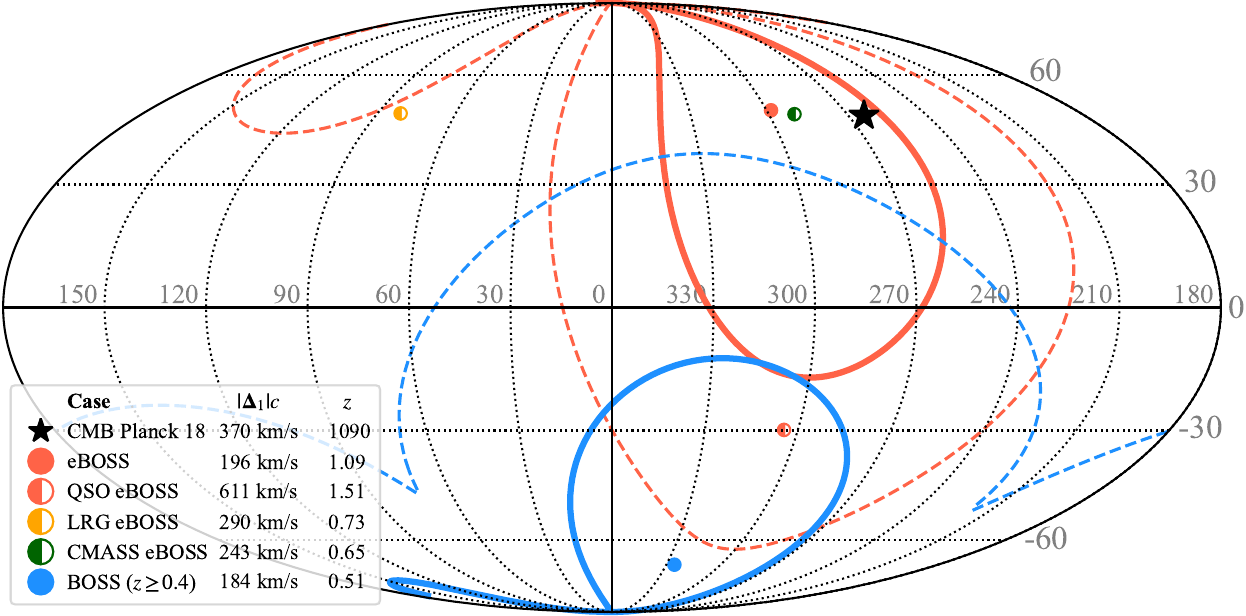}
\caption{\textbf{Final dipole measurements.} Mollweide projection displaying the $\boldsymbol{\beta}$ measurements in galactic coordinates, representing the negative of the redshift dipole ($\boldsymbol{\Delta}_{1}=-\boldsymbol{\beta}$). Contours indicate the $1\sigma$ and $2\sigma$ confidence regions for the combined measurements. The star symbol denotes the position of the CMB dipole for reference ($\boldsymbol{\Delta}_{1,\rm{CMB}}=\boldsymbol{\beta}$).}\label{fig:final_results}
\end{figure}

Our findings are depicted in Fig.~\ref{fig:final_results} and detailed in Table~\ref{tab:final_results}. The confidence contours illustrated in Fig.~\ref{fig:final_results} were generated with the \texttt{ligo.skymap} Python module\footnote{See documentation at \href{https://lscsoft.docs.ligo.org/ligo.skymap/}{lscsoft.docs.ligo.org/ligo.skymap}.}, a tool frequently utilized in the analysis of gravitational wave observations. 
Table~\ref{tab:final_results} lists the confidence intervals for both the magnitude and galactic coordinates of the velocities determined using the various catalogs and their combinations, alongside the significance of discrepancies relative to the CMB dipole. These significances were calculated for each Cartesian component and combined employing Fisher's method~\cite{Fisher1954StatisticalMethods}, see Appendix~\ref{Fisher}.

\setlength\tabcolsep{3pt}
\renewcommand{\arraystretch}{1.2}
\begin{table}
    \begin{center}
    \caption{\textbf{Dipole measurements with $1\sigma$ uncertainties and significance of tension against the CMB dipole.}}
    \label{tab:final_results}
    \begin{tabular}{c l c c c c c c c c c}
    \cmidrule{1-11}
    & & & & & \multicolumn{3}{c}{{\textbf{Dipole} $\boldsymbol{\Delta}_1$}} & & \multicolumn{2}{c}{{\textbf{Significance of tension}}} \\ 
    \cmidrule{6-8} \cmidrule{10-11}
    & \multicolumn{1}{l}{{\textbf{Case}}} & & $z$ & & $|\boldsymbol{v}|$  (km/s) & $l(^{\circ})$   &   $b(^{\circ})$ & & KDE & \texttt{Distfit}\\
     \cmidrule{1-2} \cmidrule{4-4} \cmidrule{6-8} \cmidrule{10-11}
    & CMB \T  & & 1090 &  & $369.8^{+0.1}_{-0.1}$ & $264.02^{+0.01}_{-0.01}$  & $48.253^{+0.005}_{-0.005}$ & & -- & -- \\
    \cmidrule{1-2} \cmidrule{4-4} \cmidrule{6-8} \cmidrule{10-11}
    & eBOSS \T  & & 1.09 &  & $196^{+92}_{-79}$ & $298^{+29}_{-52}$  & $50^{+8}_{-62}$ & & 2.0$\sigma$ & 2.1$\sigma$ \\
    \cmidrule{1-2} \cmidrule{4-4} \cmidrule{6-8} \cmidrule{10-11}
    & QSO eBOSS \T & & 1.51 &  & $611^{+242}_{-257}$ & $304^{+18}_{-31}$  & $-30^{+49}_{-17}$ & & 1.0$\sigma$ & 1.1$\sigma$ \\
    \raisebox{1pt}[0pt][0pt]{\rotatebox[origin=c]{90}{}} & LRG eBOSS & & 0.73 &      & $290^{+190}_{-115}$ & $85^{+31}_{-154}$  & $45^{+7}_{-74}$ & & 0.6$\sigma$ & 0.9$\sigma$ \\
    & CMASS eBOSS & & 0.65 & & $243^{+103}_{-98}$ & $301^{+14}_{-42}$  & $28^{+32}_{-39}$ & & 1.2$\sigma$ & 1.4$\sigma$ \\
    \cmidrule{1-2} \cmidrule{4-4} \cmidrule{6-8} \cmidrule{10-11}
    & BOSS ($z\geq0.4$)    & & 0.51 &  & $184^{+103}_{-80}$ & $326^{+11}_{-289}$  & $-68^{+60}_{-1}$ & & 5.3$\sigma$ & 5.6$\sigma$ \\
    \cmidrule{1-11}
    \end{tabular}
    \end{center}
    \footnotesize{
    To evaluate the two-tailed significance, we compare the dipoles to the CMB value in Cartesian coordinates using two methods: a non-parametric Kernel Density Estimation (KDE) with a Gaussian kernel and a parametric best-fit distribution analysis conducted using the \texttt{Distfit} Python module~\cite{Taskesen2020}.
    The table additionally lists the effective redshift for each scenario. For BOSS and eBOSS data, this corresponds to the weighted average redshift. 
    }
\end{table}
\renewcommand{\arraystretch}{1}

Redshift dipole measurements from the combined eBOSS data ($0.6<z<2.2$) demonstrate a $2\sigma$ tension with the CMB dipole, indicating broad agreement. Dipole measurements across individual eBOSS catalogs are also mutually consistent.
Figure \ref{fig:dipoles_diffsources} shows a very good agreement ($\sim1\sigma$ level) between these results and the ones obtained in refs.~\citealp{Ferreira:2020aqa,Horstmann:2021jjg,Darling:2022jxt,Sorrenti:2022zat}.
Our analysis using eBOSS data reveals that our motion with respect to distant galaxies and quasars is in close agreement with that suggested by the CMB dipole. This concordance implies that matter at cosmological distances of 2--5 Gpc shares the same rest frame as the CMB, located approximately 14 Gpc away.
Such findings reinforce the mathematical and philosophical foundations of our standard cosmological model, which posits large-scale homogeneity and isotropy.

However, BOSS data ($0.4<z<0.6$) suggest a deviation towards the southern hemisphere, in  strong tension ($5\sigma$) with the CMB dipole.
As previously highlighted, the BOSS mock catalogs did not incorporate realistic systematic effects, rendering the bias removal method unable to account for these unmodeled systematics, though they do incorporate veto weights to flag areas potentially affected by systematics.
While our bias correction made only minor adjustments to the observed dipoles and revealed no significant evidence of systematic effects, it is possible that the BOSS data suffer from unmodeled systematics in spectroscopic redshift determination not present in the eBOSS data. Particularly, adopting the prior $|\beta|c \leq 3200$ km/s, it was not possible to effectively minimize our dipole estimator in the redshift range $0.2<z<0.4$, corresponding to LOWZ galaxies, although the minimization worked as expected with the BOSS mocks. This discrepancy suggests that our BOSS measurement may be biased. Consequently, we will consider the eBOSS measurement as our principal result.

\begin{figure}
\centering 
\includegraphics[trim={0 0 0 0}, clip, width=\columnwidth]{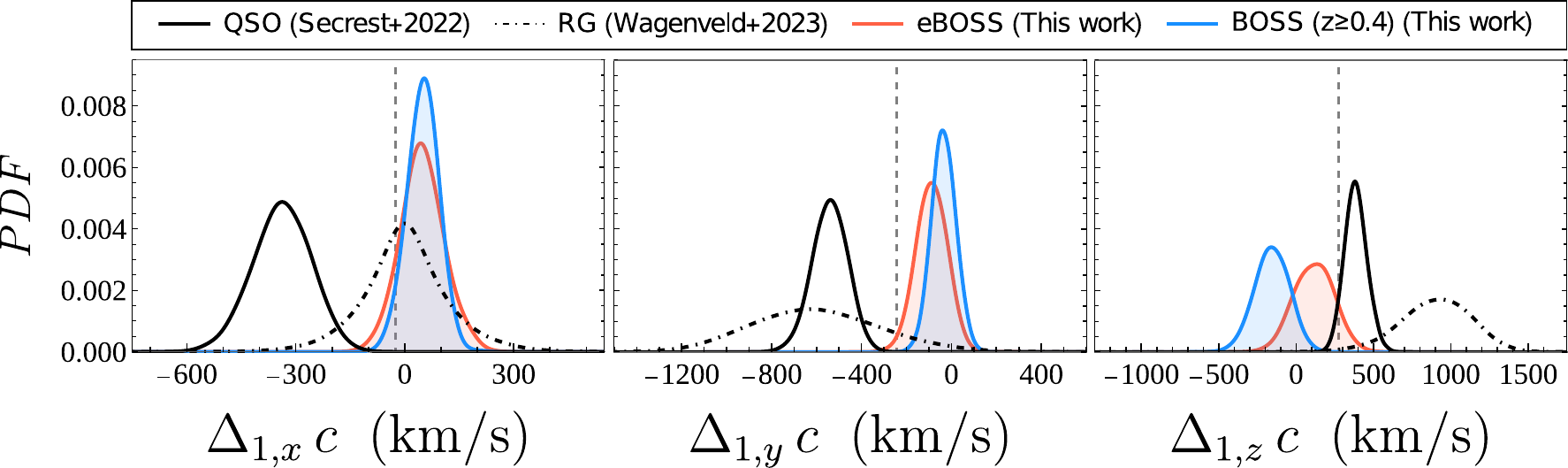}
\caption{\textbf{Comparison of dipole measurements in Cartesian coordinates with previous estimates.}}
\label{fig:compa}
\end{figure}

Recent dipole estimations from number counts of radio galaxies and quasars have reported significantly higher dipole magnitudes as compared to CMB expectations, as illustrated in Fig.~\ref{fig:dipoles_diffsources}. Specifically, ref.~\citealp{Secrest:2022uvx} reported a velocity of $750\pm 81$ km/s, and ref.~\citealp{Wagenveld:2023kvi} reported $1131\pm 156$ km/s, both indicating about 5$\sigma$ tension with the CMB estimate.
We compare our dipole measurements in Cartesian coordinates to these previous findings in Fig.~\ref{fig:compa}.\footnote{Note that the larger error bars in our longitude and latitude estimations are due to the low absolute value and our uncertainties for each Cartesian component are smaller then most of previous number count measurements, as shown in Fig.~\ref{fig:compa}.}
Considering the more trustworhty eBOSS measurement, our analysis exhibits a $3\sigma$ tension with the radio galaxy estimation of ref.~\citealp{Wagenveld:2023kvi} and a $6\sigma$ tension with the quasar measurement by ref.~\citealp{Secrest:2022uvx}, as detailed in Table~\ref{tab:compa}.
This suggests three scenarios:
\begin{enumerate}
    \item Our eBOSS redshift dipole measurement is accurate, the anomalous number count estimates are biased by systematics, and the cosmological principle holds;
    \item Our dipole measurement is biased and the anomalous number count estimates are accurate, suggesting the presence of anomalously large-scale inhomogeneities;
    \item Both measurements are accurate, indicating the presence of exotic inhomogeneities that generate density but not velocity large-scale perturbations.
\end{enumerate}
Only future data will clarify this issue.

\setlength\tabcolsep{3pt}
\renewcommand{\arraystretch}{1.2}
\begin{table}
\begin{center}
\caption{\textbf{Significance of tension of previous estimations against our results.}}
\label{tab:compa}
\begin{tabular}{lccccc}
\hline
    & \multicolumn{3}{c}{\textbf{Dipole} $\boldsymbol{\Delta}_1$} & \multicolumn{2}{c}{\bf Significance of tension with} \\ \cline{2-6} 
\textbf{Measurement}    & $|\boldsymbol{v}|$  (km/s) & $l(^{\circ})$   &   $b(^{\circ})$ & eBOSS & BOSS \\ \hline
QSO \citep[Secrest+2022,][]{Secrest:2022uvx} &  $750\pm 81$   &   $238\pm 7$   &   $31\pm 5$  &  5.6$\sigma$  & 7.6$\sigma$     \\
RG \cite[Wagenveld+2023,][]{Wagenveld:2023kvi} &  $1131 \pm 156$   &   $270\pm 10$ & $56\pm17$  &  2.7$\sigma$  &    3.9$\sigma$      \\
\hline
\end{tabular}
\end{center}
\end{table}

Progress in this area will require careful consideration of potential overlooked systematic factors in analyses based on number counts and redshifts, such as redshift evolution modeling, biases in estimators, or underestimation of uncertainties~\cite{Gibelyou:2012ri,Dalang:2021ruy,Guandalin:2022tyl,Dam:2022wwh,Peebles:2022akh,Abghari:2024eja}. Furthermore, validating the analysis pipeline using realistic mock catalogs and conducting blind data analyses, as undertaken in this study, remains critical. The incoming DESI final release data could reduce the errors by a factor of $\sim 5$, by increasing the number of tracers from $1.4$ million to $30$ millions~\cite{DESI:2016fyo}.
This enhancement is expected to significantly improve the signal-to-noise ratio and provide a definitive examination of the reference frame throughout cosmic history. However, this study has revealed that subtle redshift systematics, which may be negligible for BAO analyses, can significantly affect the measurement of the dipolar modulation in redshift caused by our relative velocity. Therefore, a robust determination of this effect will require meticulous characterization of the redshift errors in spectroscopic surveys like DESI.


\acknowledgments

\small{
It is a pleasure to thank Nathan Secrest, Subir Sarkar, Tomáš Šoltinský, Christophe Yèche, Pierre Fleury and Julián Bautista for useful comments and discussions. 
\textbf{Funding:} PSF thanks FAPES (Brazil) for financial support. VM thanks CNPq (Brazil) and FAPES (Brazil) for partial financial support.
\textbf{Sci-Com cluster:} We acknowledge the use of the computational resources provided by the Sci-Com Lab of the Department of Physics at UFES, which was funded by FAPES.   
\textbf{Joint CHE / Milliways cluster:} We acknowledge the use of the computational resources of the joint CHE / Milliways cluster, supported by a FAPERJ grant E-26/210.130/2023.
\textbf{HOTCAT cluster: } We acknowledge the use of the HOTCAT computing infrastructure of the Astronomical Observatory of Trieste of the National Institute for Astrophysics (INAF, Italy).
\textbf{MultiDark-Patchy mocks:} The massive production of all MultiDark-Patchy mocks for the BOSS Final Data Release has been performed at the BSC Marenostrum supercomputer, the Hydra cluster at the Instituto de Fısica Teorica UAM/CSIC, and NERSC at the Lawrence Berkeley National Laboratory. We acknowledge support from the Spanish MICINNs Consolider-Ingenio 2010 Programme under grant MultiDark CSD2009-00064, MINECO Centro de Excelencia Severo Ochoa Programme under grant SEV- 2012-0249, and grant AYA2014-60641-C2-1-P. The MultiDark-Patchy mocks was an effort led from the IFT UAM-CSIC by F. Prada's group (C.-H. Chuang, S. Rodriguez-Torres and C. Scoccola) in collaboration with C. Zhao (Tsinghua U.), F.-S. Kitaura (AIP), A. Klypin (NMSU), G. Yepes (UAM), and the BOSS galaxy clustering working group.
\textbf{EZmocks:}  We acknowledge the use of the EZmocks galaxy mock catalogues with redshift evolution and systematics for galaxies and quasars of the final SDSS-IV data release \cite{Zhao:2020bib}.
\textbf{SDSS-III (BOSS) data:} Funding for SDSS-III has been provided by the Alfred P. Sloan Foundation, the Participating Institutions, the National Science Foundation, and the U.S. Department of Energy Office of Science. The SDSS-III web site is http://www.sdss3.org/. SDSS-III is managed by the Astrophysical Research Consortium for the Participating Institutions of the SDSS-III Collaboration including the University of Arizona, the Brazilian Participation Group, Brookhaven National Laboratory, Carnegie Mellon University, University of Florida, the French Participation Group, the German Participation Group, Harvard University, the Instituto de Astrofisica de Canarias, the Michigan State/Notre Dame/JINA Participation Group, Johns Hopkins University, Lawrence Berkeley National Laboratory, Max Planck Institute for Astrophysics, Max Planck Institute for Extraterrestrial Physics, New Mexico State University, New York University, Ohio State University, Pennsylvania State University, University of Portsmouth, Princeton University, the Spanish Participation Group, University of Tokyo, University of Utah, Vanderbilt University, University of Virginia, University of Washington, and Yale University. 
\textbf{SDSS-IV (eBOSS) data:} Funding for the Sloan Digital Sky Survey IV has been provided by the Alfred P. Sloan Foundation, the U.S. Department of Energy Office of Science, and the Participating Institutions. SDSS acknowledges support and resources from the Center for High-Performance Computing at the University of Utah. The SDSS web site is www.sdss4.org. SDSS is managed by the Astrophysical Research Consortium for the Participating Institutions of the SDSS Collaboration including the Brazilian Participation Group, the Carnegie Institution for Science, Carnegie Mellon University, Center for Astrophysics | Harvard \& Smithsonian (CfA), the Chilean Participation Group, the French Participation Group, Instituto de Astrofísica de Canarias, The Johns Hopkins University, Kavli Institute for the Physics and Mathematics of the Universe (IPMU) / University of Tokyo, the Korean Participation Group, Lawrence Berkeley National Laboratory, Leibniz Institut für Astrophysik Potsdam (AIP), Max-Planck-Institut für Astronomie (MPIA Heidelberg), Max-Planck-Institut für Astrophysik (MPA Garching), Max-Planck-Institut für Extraterrestrische Physik (MPE), National Astronomical Observatories of China, New Mexico State University, New York University, University of Notre Dame, Observatório Nacional / MCTI, The Ohio State University, Pennsylvania State University, Shanghai Astronomical Observatory, United Kingdom Participation Group, Universidad Nacional Autónoma de México, University of Arizona, University of Colorado Boulder, University of Oxford, University of Portsmouth, University of Utah, University of Virginia, University of Washington, University of Wisconsin, Vanderbilt University, and Yale University.

\paragraph{Data availability}
All data is available starting from \href{https://data.sdss.org/sas/}{data.sdss.org/sas}.
The SDSS DR16 data used to perform the eBOSS measurements is available at \texttt{dr17/eboss/lss/catalogs/DR16}, and their respective EZmock catalogs at \texttt{dr17/eboss/lss/EZmocks/v1\_0\_0/realistic}. The SDSS DR12 BOSS data (CMASSLOWZTOT) is accessible at \texttt{dr12/boss/lss}, and their respective MultiDark-Patchy mocks at \texttt{dr12/boss/lss/dr12\_multidark\_patchy\_mocks}.

\paragraph{Code availability}
The python code of the estimator and mock generation is hosted at\\
\href{https://www.github.com/pdsferreira/tomographic-redshift-dipole}{github.com/pdsferreira/tomographic-redshift-dipole}.

\paragraph{Author contributions}
PSF and VM conceptualized the study and devised the methodology. PSF was responsible for coding and conducting the data analysis. Both PSF and VM collaboratively authored the manuscript.

\appendix
\section{Pipeline}
\label{pipe}

Figure~\ref{fig:pipeline} illustrates the comprehensive pipeline developed for this study.

\begin{figure}[h!]
\centering 
\includegraphics[trim={0 0 0 0}, clip, width=.7\columnwidth]{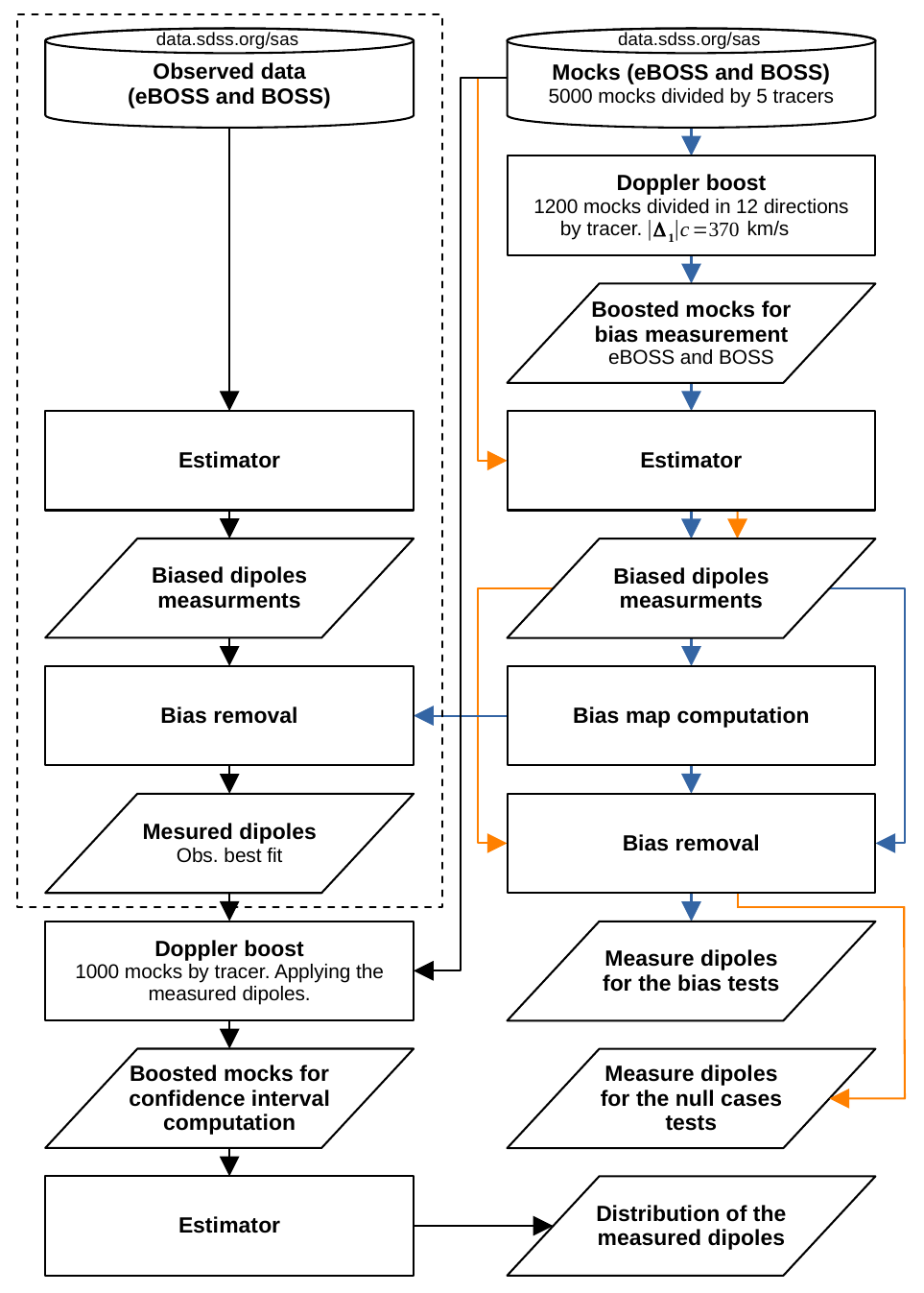}
\caption{\textbf{Complete pipeline used in this work.} The dashed region represents the steps using observed data. Orange lines indicate the steps of the null case tests (without boost), while blue lines indicate the steps of the bias measurement.}
\label{fig:pipeline}
\end{figure}

\section{Fisher's method}
\label{Fisher}

Fisher's method is utilized to consolidate results from multiple independent tests that assess the same  hypothesis~\cite{Fisher1954StatisticalMethods}. It aggregates the $p$-values from each test into a single test statistic, $X^2_{2k}$, through the equation:
\begin{equation}
X^2_{2k} = -2 \sum_{i=1}^k \ln p_i,
\end{equation}
where $p_i$ represents the $p$-value for the $i$-th hypothesis test. Assuming the null hypothesis is valid and the $p_i$ values (or their corresponding test statistics) are independent, $X^2_{2k}$ follows a chi-squared distribution with $2k$ degrees of freedom, $k$ being the count of tests being combined. This can be used to determine the combined $p$-value of all Cartesian components of the measured dipole vector, which we express in $\sigma$ units.

\section{Redshift error evolution}
\label{revo}

We examined in Fig.~\ref{fig:zerror_evo} the potential evolution of redshift errors with increasing redshift, confirming that a bin width of 0.003 is sufficient across the entire redshift range.

\begin{figure}[h!]
\centering 
\includegraphics[trim={0 0 0 0}, clip, width=1\columnwidth]{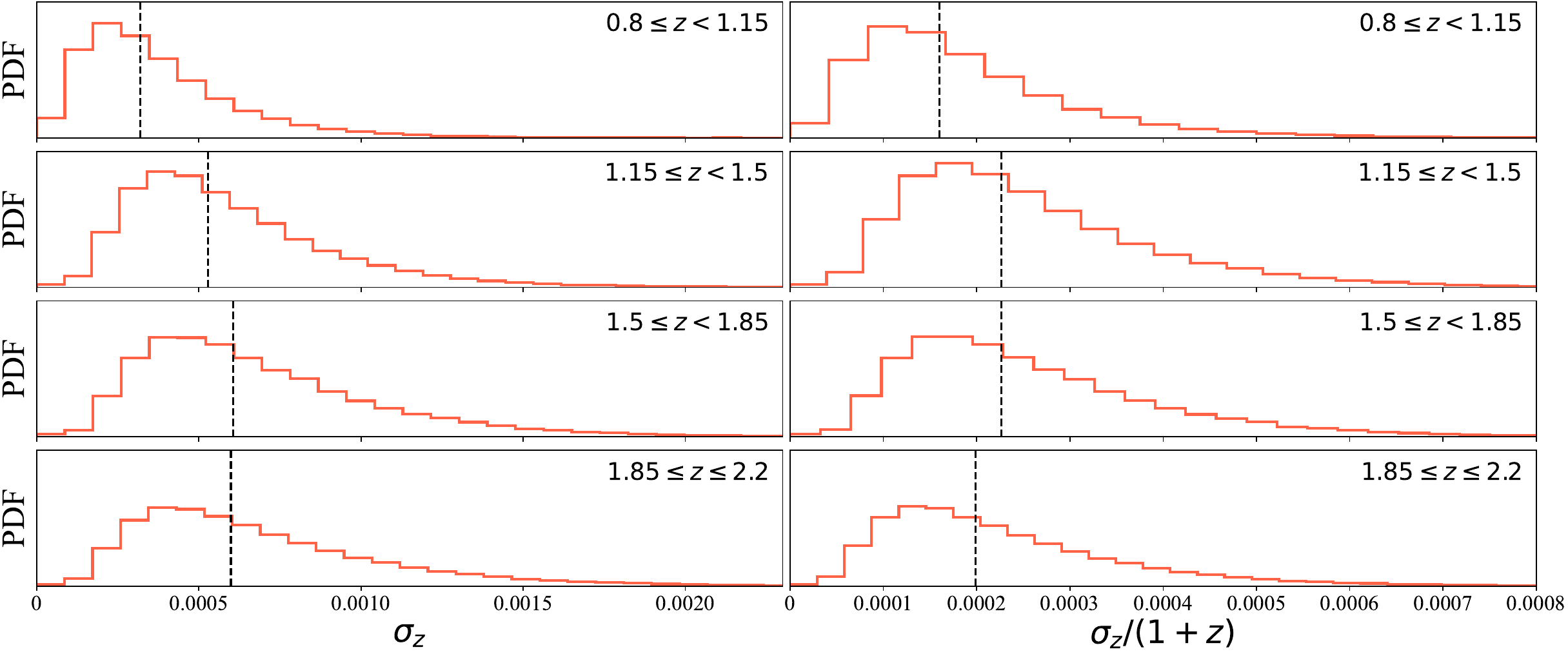}
\caption{\textbf{Redshift error evolution of the QSO eBOSS sample.} Dashed lines mark the median error.}
\label{fig:zerror_evo}
\end{figure}

\newpage

\bibliographystyle{JHEP}
\bibliography{biblio.bib}

\end{document}